\documentclass{mem}
\usepackage{subfigure}
\usepackage{natbib}\usepackage{txfonts}\usepackage{balance}
\usepackage{graphicx}
\usepackage[a4paper]{hyperref}
\idline{75}{282}
\begin{document}
\def\teff{$T\rm_{eff }$}
\def\kms{$\mathrm {km s}^{-1}$}

\title{Constraints on the Cosmological parameters by means of the clusters mass function.
}

   \subtitle{}

\author{
A. \, Del Popolo\inst{1} 
          }

  \offprints{A. Del Popolo: antonino.delpopolo@oact.inaf.it}

\institute{
Dipartimento di Fisica e Astronomia -- Universit\'a di Catania, Viale Andrea Doria 6, I-95125, Catania, Italy\\
\email{adelpopolo@astrct.oact.inaf.it}
}

\authorrunning{Del Popolo}

\titlerunning{Cosmological parameters constraints}

\abstract{

We present constraints on the values of $\Omega_m$, $n$, $\sigma_8$, obtained from measurements of the X-ray luminosity function of galaxy clusters as compiled in EMSS, RDCS and BCS galaxy cluster samples. The values obtained $\Omega_m=0.25^{+0.15}_{-0.05}$, $n=-1^{+0.05}_{-0.05}$, and
$\sigma_8=0.8^{+0.1}_{-0.1}$ are in agreement with WMAP data.
We then put constraints on the quoted parameters and the dark-energy equation-of-state parameter, $w$,
using Chandra observations of 37 clusters with $<z>$=0.55 derived from 400 deg$^2$ ROSAT serendipitous survey and 49 brightest $z \simeq 0.05$ clusters detected in the All-Sky Survey. In the case of $\Lambda$CDM model, we get 
$\Omega_m=0.25^{+0.1}_{-0.08}$ and $\sigma_8=0.75^{+0.08}_{-0.05}$,
while  in the case of the $w$CDM model, we get 
$\Omega_m=0.28^{+0.10}_{-0.10}$, $w=-1.25^{+0.30}_{-0.35}$, and $\sigma_8=0.8^{+0.09}_{-0.09}$.  
Our results are consistent with those from recent analyses of type Ia supernovae, cosmic microwave background anisotropies, the X-ray gas mass fraction of relaxed galaxy clusters, baryon acoustic oscillations and cosmic shear. The improvement in data quality from EMSS, RDCS, and BCS to Chandra observations leads to an improvement in the constraints even if not of the same
entity of the improvement in data.
\keywords{Cosmological parameters - large-scale structure of Universe - X-rays: galaxies: clusters
}
}
\maketitle{}

\section{Introduction}
A cluster of galaxies is a large collection of galaxies held together by their mutual
gravitational attraction. 
%
%
The galaxy clusters occupying the
high-mass tail of this population provide a powerful and
relatively clean tool for cosmology, since their growth is
predominantly determined by linear gravitational processes, 
their number density is exponentially sensitive to the size of the perturbations, and hence can strongly
constrain it.
Starting with 1990's, they have consistently indicated low values of 
$\Omega_m$ (both from the baryonic fraction arguments (White et al. 1993) and measurements of the evolution 
in the cluster number density (Eke et al. 1998; Borgani et al. 2001)) and low values of
$\sigma_8$\footnote{$\sigma_8$ is the amplitude of the mass density fluctuation power spectrum over spheres of radius $8 h^{-1} {\rm Mpc}$, and $M_8$ 
is the mean mass within these spheres} (Henry \& Arnaud 1991; Reiprich \& B\"oringer 2002; Schuecker et al. 2003) --a result since then confirmed by cosmic microwave background (CMB) studies, cosmic
shear, and other experiments (Spergel et al. 2007; 
Fu et al. 2008). 
For precision's sake, cluster surveys in the local universe are particularly useful
for constraining a combination of the matter density parameter 
$\Omega_m$ and $\sigma_8$.
%
Surveys that probe
the cluster population at higher redshift are sensitive to the growth of density fluctuations, allowing
one to break the $\Omega_m$-$\sigma_8$ degeneracy that arises from local cluster abundance constraints (Eke, Cole \& Frenk 1996; Bahcall \& Fang 1998).
%
%
Recently, X-ray study of the evolution of the cluster
mass function at $z =$ 0-0.8 have convincingly demonstrated that the growth of cosmic
structure has slowed down at $z < 1$ due to the effects of dark energy, and these measurements
have been used to improve the determination of the equation of state parameter (Vikhlinin et al. 2008).

The recipe of constraining cosmological parameters by means of clusters is composed of three ingredients:
1) The predicted mass function of clusters, $n(M,z)$, as a function of cosmological parameters ($\sigma_8$, $\Omega_M$, $w_0$, $w_a$, etc)\footnote{$w_0$ and $w_a$ represents the dark-energy equation-of-state parameter, $w$ at $z=0$ and at generic redshift $z$, or expansion parameter $a$, respectively.}. The mass function can be calculated analytically (Press \& Schechter 1974; Sheth \& Tormen 2002; Del Popolo 2006) or fitting the results of numerical simulations (Jenkins et al. 2001; Reed et al. 2003, Yahagi et al. 2004; Tinker et al. 2008). \\
2) Sky surveys with well understood selection functions to find clusters, as well as a
relation linking cluster mass with an observable. A successful solution to the former requirement has been to identify
clusters by the X-ray emission produced by hot intracluster gas, notably using data from the ROSAT All-Sky Survey (RASS; Tr\"umper 1993). The ROSAT Brightest Cluster Sample (BCS; Ebeling et al. 1998, 2000) and ROSAT-ESO Flux Limited X-ray sample (REFLEX; B\"ohringer et al. 2004) together cover approximately two-thirds of the sky out to redshift $z \simeq 0.3$ and contain more than 750 clusters. The Massive Cluster Survey (MACS; Ebeling et al. 2001, 2007)–- which contains 126 clusters and covers 55 per cent of the sky – extends these data to $z \simeq 0.7$.
ROSAT 400 sq. degree survey (Vikhlinin et al '08), serendipitous cluster catalogue containing 266 groups/clusters.
3) A tight, well-determined scaling relation between survey observable (e.g. $L_x$) and mass, with minimal intrinsic scatter.

The most straightforward mass–-observable relation to complement these X-ray flux-limited surveys is the mass–-X--ray luminosity relation. For sufficiently massive (hot) objects at the relevant redshifts, the conversion from X-ray flux to luminosity is approximately independent of temperature, in which case the luminosities can be estimated directly from the survey flux and the selection function is
identical to the requirement of detection. 
A disadvantage is that there is a large scatter in cluster luminosities at fixed mass; however, sufficient data allow this scatter to be quantified
empirically. Alternative approaches use cluster temperature (Henry 2000; Seljak 2002; Pierpaoli et al. 2003; 
Henry 2004),
gas fraction (Voevodkin \& Vikhlinin 2004) or $Y_X$ parameter (Kravtsov et al. 2006) to achieve tighter mass–observable relations at the expense of reducing the size of the samples available for analysis. The need to quantify the selection
function in terms of both X-ray flux and a second observable additionally complicates these efforts.
In this paper, we use the observed X-ray luminosity
function to investigate two cosmological scenarios, assuming
a spatially flat metric in both cases: the first includes dark
energy in the form of a cosmological constant ($\Lambda$CDM); the
second has dark energy with a constant equation-of-state
parameter, $w$ ($w$CDM). 
The theoretical background for this work is reviewed in Section 2. Section 3 presents the results and Section 4 the
conclusions.

\begin{figure}[t!]
\begin{minipage}[b]{0.9\linewidth} 
\centering
\resizebox{\hsize}{!}{\includegraphics[clip=true]{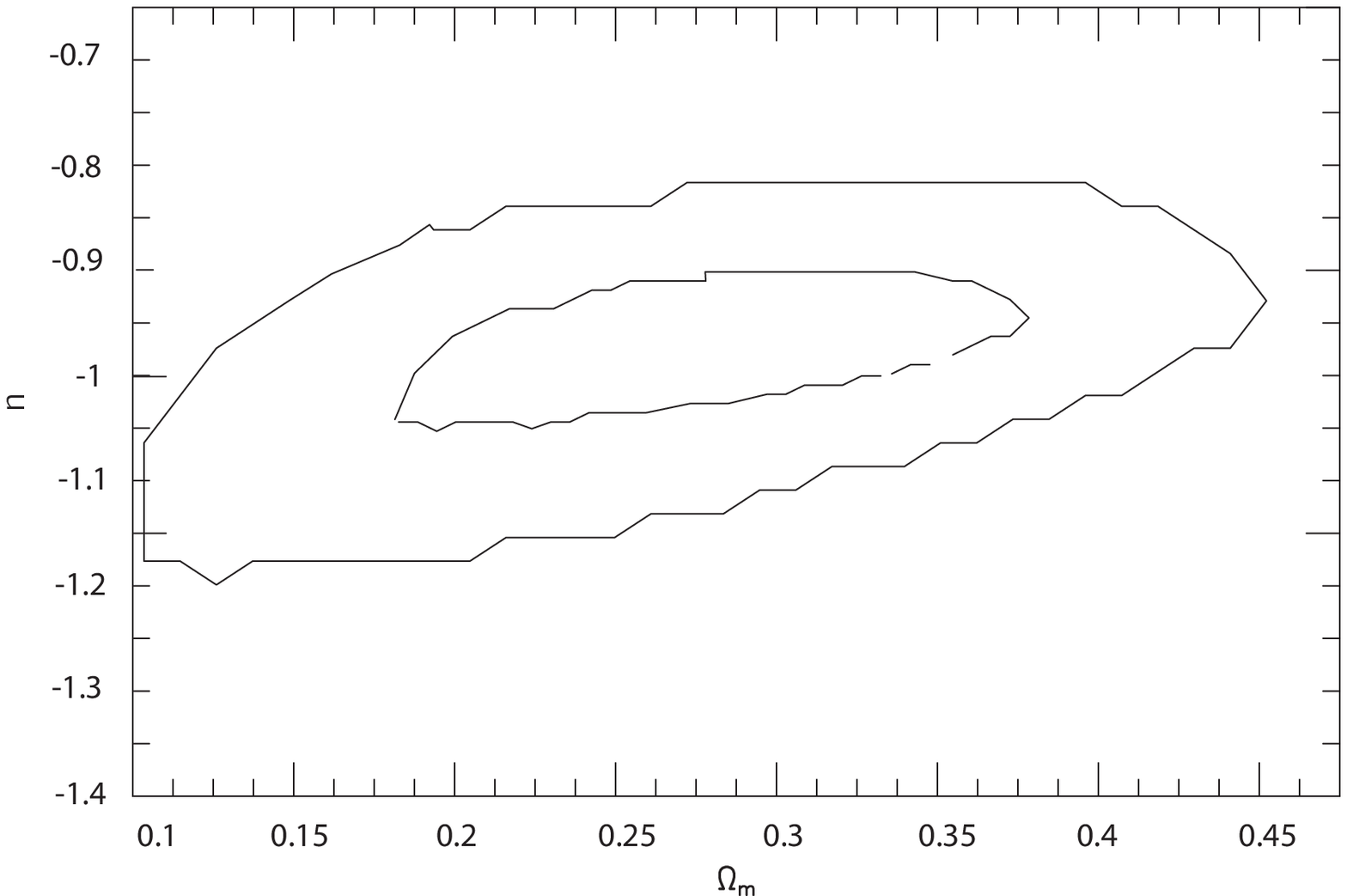}}
\resizebox{\hsize}{!}{\includegraphics[clip=true]{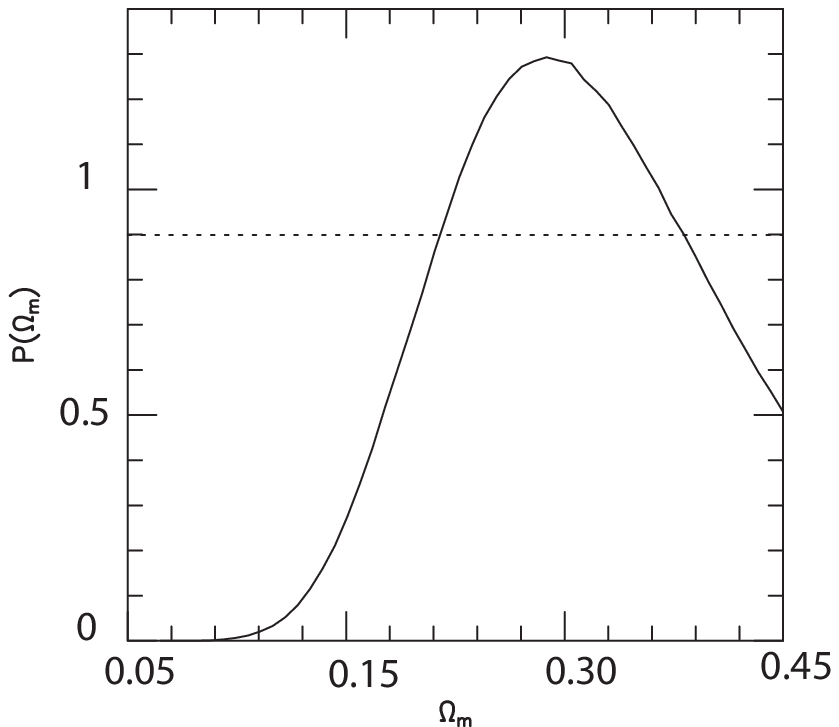}}
\resizebox{\hsize}{!}{\includegraphics[clip=true]{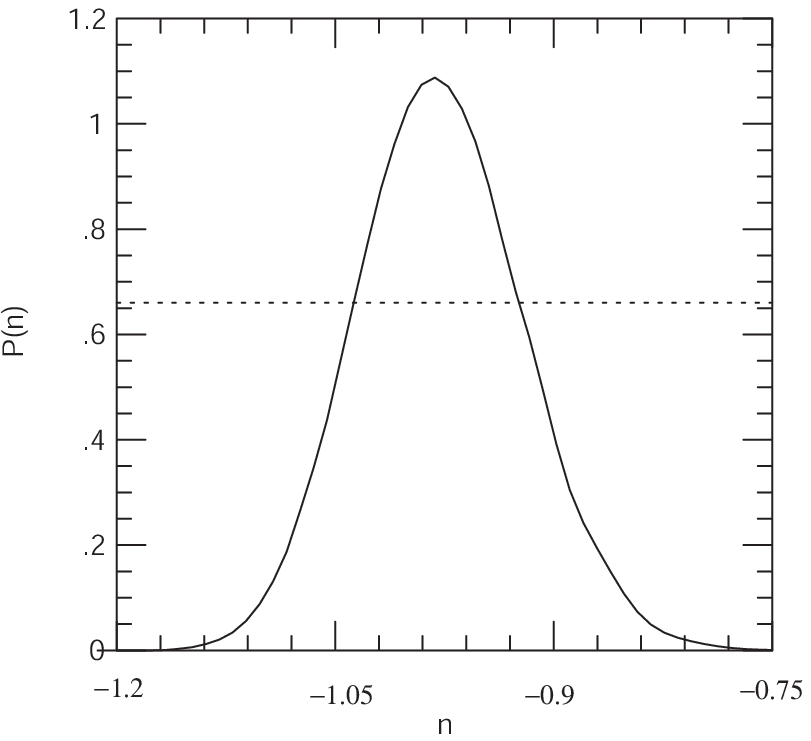}}
\caption{\footnotesize
(a) The 1 and 3 sigma credible regions of the marginalized posterior probability  distribution P(Omega,n) (b) Marginalized posterior probability 
distribution P($\Omega_m$). The dotted line is the 3 $\sigma$ confidence interval (c) Marginalized posterior probability 
Distribution P(n). The dotted line is the 3 $\sigma$ confidence interval. 
}
\end{minipage}
\label{eta}
\end{figure}
%

%
\begin{figure}[t!]
\begin{minipage}[b]{0.75\linewidth} 
\centering
\resizebox{\hsize}{!}{\includegraphics[clip=true]{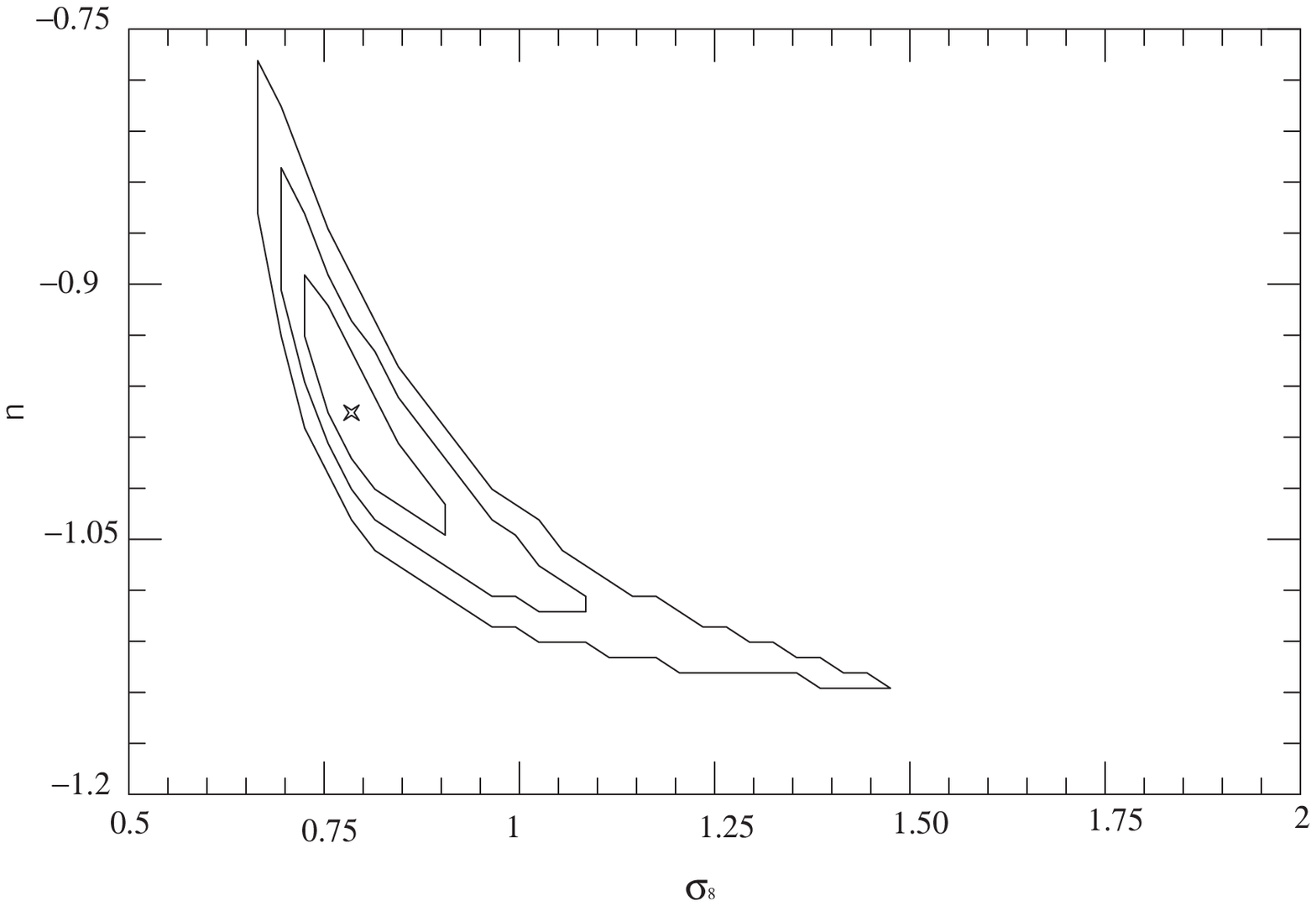}}
\resizebox{\hsize}{!}{\includegraphics[clip=true]{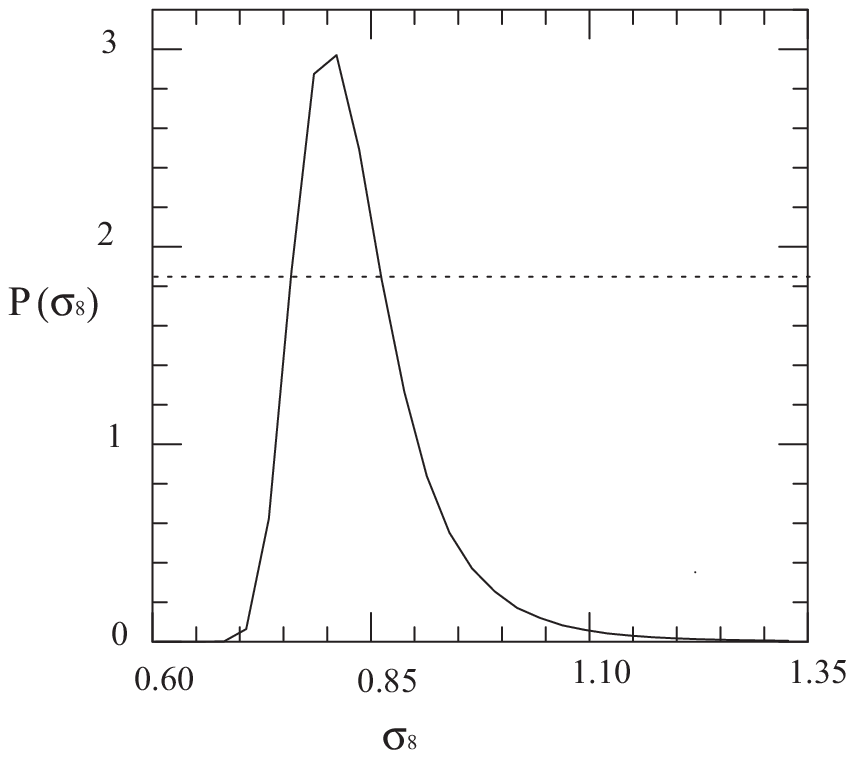}}
\caption{\footnotesize
(a) The 1, 2  and 3 sigma credible regions of the marginalized posterior probability distribution P(n, $\sigma_8$). 
(b) Marginalized posterior probability  distribution P($\sigma_8$). The dotted line is the 3 $\sigma$ confidence interval.
}
\end{minipage}
\label{eta}
\end{figure}
%


%
\begin{figure}[t!]
\begin{minipage}[b]{0.7\linewidth} 
\centering
\resizebox{\hsize}{!}{\includegraphics[clip=true]{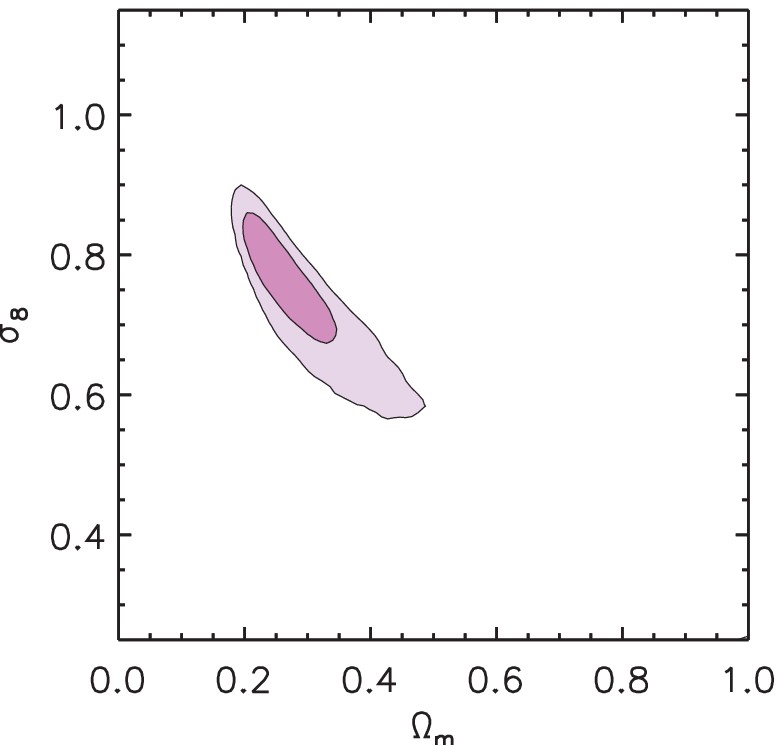}}
\resizebox{\hsize}{!}{\includegraphics[clip=true]{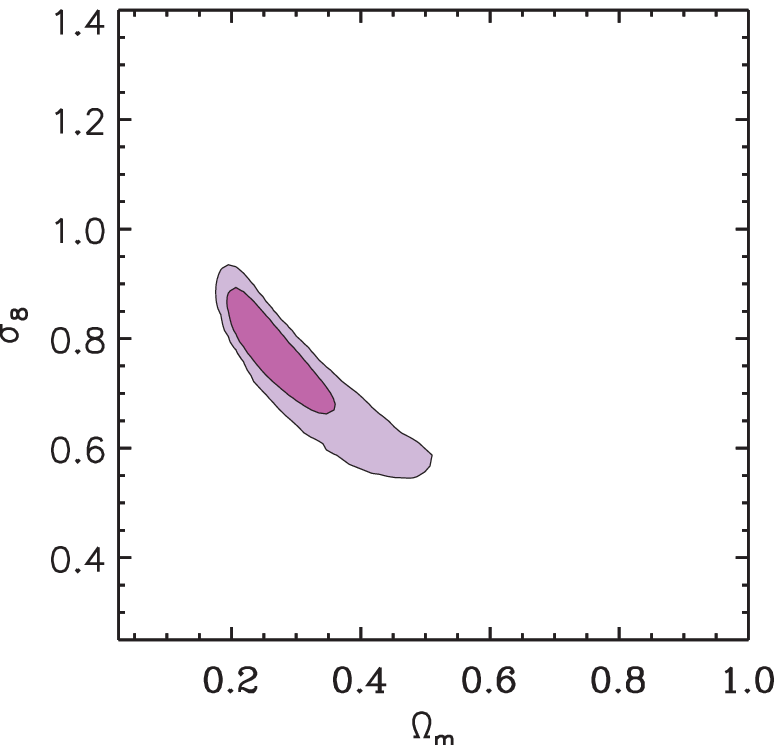}}
\resizebox{\hsize}{!}{\includegraphics[clip=true]{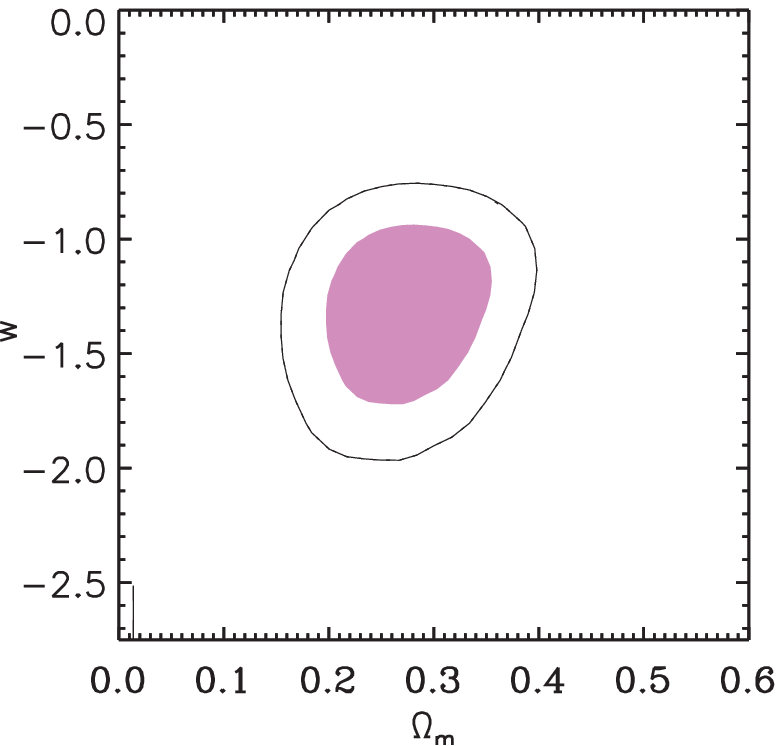}}
\caption{\footnotesize
(a) The 68\% 
and 95\% 
credible regions of the marginalized posterior probability distribution P($\sigma_8$,$\Omega_m$) for a $\Lambda$CDM  model. 
(b) Same as panel (a) but for a $w$CDM model. 
(c) The 68\% 
and 95\% 
credible regions of the marginalized posterior probability distribution P(w,$\Omega_m$) for a 
$w$CDM  model.
}
\end{minipage}
\label{eta}
\end{figure}
%


\section{Theory}

%
In this section, 
We derive an expression for the XLF (using now the mass function and M-T, L-T relations obtained in Del Popolo (2002), and Del Popolo el al. (2005)
respectively) and then we get some constraints to $\Omega_{\rm m}$ and $n$, by using {\rm EMSS}, the {\it ROSAT} BCS, the RDCS (Borgani et al. 2001) samples.  

The mass function was calculated according to the model of Del Popolo (2006)   
in agreement with N-body simulations like those of Yahagi, Nagashima \& Yoshii 2004).

In order to convert the mass function in a Luminosity function (XLF), first we convert mass into intra-cluster gas temperature, by means of the $M-T_{x}$ relation, and then the temperature is converted into X-ray luminosity, by means of the $L_{x}-T_{x}$ relation. 
The $M-T_{x}$ relation is obtained analytically using the model of Del Popolo (2002),  
while the $L_{x}-T_{x}$ relation is that obtained in Del Popolo, Hiotelis \& Pe\'narrubia (2005) based on an improvement of the  
Punctuated Equilibrium Model (PEM) of Cavaliere et al. (1997, 1998, 1999) 
%
%
We performed the calculations assuming 
$\Lambda$CDM spectrum $P(k) \propto k^{n} T^2(k)$, where the transfer function $T(k)$ 
is given by, BBKS (equation~(G3)). 
%
The total number of X-ray clusters observed between luminosity and redshift limits $L_{\rm l}<L<L_{\rm u}$ and
$z_{\rm l}<z<z_{\rm u}$, i.e. the cumulative luminosity function is given by:
\begin{eqnarray}
N(L_{\rm l}, L_{\rm u}, z_{\rm l}, z_{\rm u})&=& \int_{L_{\rm l}}^{L_{\rm u}}\int_{z_{\rm l}}^{z_{\rm u}}
A(L,z) \nonumber\\
& &
\frac{dn (L_1,z)}{dL} d L dV(z)
\label{eq:cumul}
\end{eqnarray}
where $A(L_1,z)$ is the area of the sky that an X-ray survey samples at redshift $z$ as a function of luminosity $L$.
%
%
The previous model, 
consists of the parameters: $h$, $\Omega_b h^2$, $\Omega_c h^2$$^,$\footnote{$\Omega_b$ and $\Omega_c$ are the baryon and CDM densities ($\Omega_m=\Omega_b+\Omega_c$)}, $\sigma_8$, n, $w$, $A_1$. We use priors as in Mantz et al. (2007) for $h$, $\Omega_b h^2$, $n$, $A_1$, 
and we obtain constraints for $\Omega_{\rm m}$, $n$, $\sigma_8$, and $w$ comparing model and data through a Bayesisn inference analysis.
In order to perform the quoted analysis, we need
the posterior probability distribution for $\Omega_{\rm m}$, $n$, $\sigma_8$, namely $P(\Omega_{\rm m},n,\sigma_8)$
which is obtained by normalizing the product of the prior probability distribution and the likelihood function (Gregory \& Loredo 1992). 
In order to constraint, $\Omega_{\rm m}$, $n$ and $\sigma_8$, we use the likelihood function $L(\Omega_{\rm m},n,\sigma_8)$ (see Cash 1979, Reichart et al. 1999):
\begin{equation}
L(\Omega_{\rm m},n,\sigma_8)=\prod_{i\equiv 1}^{N_{tot}}P(L_{i},z_{i}|\Omega _{m},n,\sigma_8)
\label{eq:likel}
\end{equation}
where $P(L_{i},z_{i}|\Omega _{m},n,\sigma_8)$ is the probability that the $i$th X-ray cluster fits our model, given values of
$\Omega _{\rm m}$, $n$, $\sigma_8$. For our model, this probability is given by (e.g., Cash 1979):
\begin{equation}
P(L,z|\Omega _{m},n,\sigma_8)=
\frac{A(L_1,z) \frac{dn_{\rm c} (L,z)}{dL} \frac{dV(z)}{dz}}{N(L_{\rm l}, L_{\rm u}, z_{\rm l}, z,_{\rm u})}
\label{eq:prob}
\end{equation}
In Eq. (\ref{eq:likel}), $N_{\rm tot}$ is the total number of X-ray clusters in the same region of the $L-z$ plane as that over which $N(L_{\rm l}, L_{\rm u}, z_{\rm l}, z,_{\rm u})$ is defined. 
%
The posterior probability distribution for one of the parameters, e.g., $P(\Omega_{\rm m})$, 
is obtained by marginalizing the posterior probability distribution for all the three parameters, 
$P(\Omega_{\rm m},n,\sigma_8)$, over the other parameters (Gregory \& Loredo 1992). 
The credible regions (1, 2, 3 $\sigma$) are determined by integrating the posterior probability distribution over 
the most probable region of its parameter space until 68,3\%, 95.4\% and 99.73\% (respectively) of this distribution has been integrated.

\section{Results}

The constraints obtained from the comparison of the model of the previous section with EMSS, RDCS, and BCS are shown in Fig. 1-Fig. 2.
In Fig. 1a, we plot the 1 and 3 $\sigma$ confidence regions of the marginalized posterior probability 
distribution P($\Omega_m$,n) for a $\Lambda$CDM
power spectrum,
while in Fig. 1b, 1c, the marginalized posterior probability distribution P($\Omega_m$), 
and the marginalized posterior probability distribution P(n). The dotted line represents the 3 $\sigma$ confidence interval. Fig. 1 shows that 
$\Omega_m=0.25^{+0.15}_{-0.05}$ and $n=-1^{+0.05}_{-0.05}$.
Similarly, in Fig. 2a, we plot the 1, 2  and 3 $\sigma$ credible regions of  the marginalized posterior probability distribution P(n, $\sigma_8$), while in Fig. 2b the marginalized posterior probability  distribution P($\sigma_8$). As in Fig. 1, the dotted line represents the 
3 $\sigma$ confidence interval. The constraint that we obtain for $\sigma_8$ is  $\sigma_8=0.8^{+0.1}_{-0.1}$.
Using the same model, we have performed the same analysis for the $\Lambda$CDM and $w$CDM models
using more recent data, namely Chandra observations of 37 clusters with $<z>$=0.55 derived from 400 deg$^2$ ROSAT serendipitous survey and 49 brightest $z \simeq 0.05$ clusters detected in the All-Sky Survey.
Fig. 3a plots the 68.3 and 95.4 \% confidence constraints on $\sigma_8$ and $\Omega_m$ for the $\Lambda$CDM model. The constraints from the 
cluster sample used are $\Omega_m=0.25^{+0.1}_{-0.08}$ and $\sigma_8=0.75^{+0.08}_{-0.05}$.
These constraints are in good agreement with recent independent results from the CMB (Spergel et al. 2007) and cosmic shear, as measured in the 100 Square Degree survey (Benjamin et al. 2007) (Figure 8) and CFHTLS Wide field (Fu et al. 2008). Our results are also in good
overall agreement with previous findings based on the observed X-ray luminosity and temperature functions of clusters (e.g. Eke et al. 1998; 
Schuecker et al. 2003; Henry 2004),
Our result on $\Omega_m$ is in excellent agreement with current constraints based on cluster $f_{\rm gas}$ data
(Allen et al. 2008 and references therein) and the power spectrum of galaxies in the 2dF galaxy redshift survey (Cole et al. 2005) and Sloan Digital Sky Survey (SDSS) (Eisenstein et al. 2005; Tegmark et al. 2006; Percival et al. 2007), as well as the combination of CMB data with a variety
of external constraints (Spergel et al. 2007). 
%
%

In the case of the $w$CDM model, Fig. 3a shows the joint constraints on $\Omega_m$ and $\sigma_8$ from the luminosity function data, while Fig. 3b 
displays constraints on $\Omega_m$ and $w$. The marginalized results from the X-ray luminosity function data are 
$\Omega_m=0.28^{+0.10}_{-0.10}$, $w=-1.25^{+0.30}_{-0.35}$, and $\sigma_8=0.8^{+0.09}_{-0.09}$.  
XLF results are consistent with each WMAP, SNIa data, and cluster $f_{\rm gas}$ data, and with the cosmological-constant
model ($w = -1$). 
The results for $\sigma_8$ presented here are somewhat different
than those from previous work based on the BCS and REFLEX data.
The magnitude of this discrepancy underscores the need for an improved understanding of the observational biases resulting from 
asphericity, projection effects and hydrostatic disequilibrium. 
More advanced and comprehensive simulations, calibrated by gravitational lensing studies, show considerable promise
in this area. 
With regard to dark energy, the systematics due to the theoretical mass function and the redshift evolution of the
mass–-observable relation are also important. 
The former can be addressed with a large suite of cosmological simulations, The latter 
necessitates a more rigorous study of galaxy cluster virial relations and their evolution

\section{Conclusions}

In this paper, we presented constraints on the values of $\Omega_m$, $n$, $\sigma_8$, and $w$ 
obtained from measurements of the X-ray luminosity function of galaxy clusters as compiled in EMSS, RDCS and BCS galaxy cluster samples. 
The values obtained using EMSS, RDCS, and BCS data are $\Omega_m=0.25^{+0.15}_{-0.05}$, and $n=-1^{+0.05}_{-0.05}$
$\sigma_8=0.8^{+0.1}_{-0.1}$, in agreement with WMAP data.
Using Chandra observations of clusters from 400 deg$^2$ ROSAT serendipitous survey, we get $\Omega_m=0.25^{+0.1}_{-0.08}$ and $\sigma_8=0.75^{+0.08}_{-0.05}$, for a $\Lambda$CDM model, while  in the case of the $w$CDM model, we get 
$\Omega_m=0.28^{+0.10}_{-0.10}$, $w=-1.25^{+0.30}_{-0.35}$, and $\sigma_8=0.8^{+0.09}_{-0.09}$, in agreement with 
analyses of type Ia supernovae, cosmic microwave background anisotropies, the X-ray gas mass fraction of relaxed galaxy clusters, baryon acoustic oscillations and cosmic shear. 
\begin{acknowledgements}
I am grateful to thank Franco Giovanelli for inviting me to Frascati Workshop 2009.
\end{acknowledgements}

\bibliographystyle{aa}

\noindent {\bf DISCUSSION}

\noindent {\bf G. S. BISNOVATYI-KOGAN:} What about the acoustic oscillations in the mass function: can you reveal them? \\
\noindent {\bf A. DEL POPOLO:} 
Baryon Scoustic Oscillations (BAOs) are mostly viewed in the two-point redshift-space correlation function and its
Fourier transform (the spectrum). Eisenstein (2005) detected the acoustic peak in the SDSS LRG survey, while Cole et al. (2005) discovered them in  the 2DF spectrum. BAOs detected in 2dFGRS and SDSS have been used to constrain cosmological models by Percival et al. (2007). \\
\noindent {\bf J. BECKMAN:} Am I correct in seeing that the value of $n$ in your results
is not consistent with the value 1? \\
\noindent {\bf A. DEL POPOLO:} 
The results of the model using older data (EMSS, RDCS, BCS) are consistent with the value 1 ($n=-1^{+0.05}_{-0.05}$).
\end{document}